# Effect of W and Mo co-doping on the photo- and thermally stimulated luminescence and defects creation processes in $Gd_3(Ga,Al)_5O_{12}$:Ce crystals

S. Zazubovich,[1] V. Laguta,[2] K. Kamada,[3] A. Yoshikawa,[3] K. Jurek,[2] M. Nikl[2]

[1]*Institute of Physics, University of Tartu, W. Ostwaldi 1, 50411 Tartu, Estonia*
[2]*Institute of Physics AS CR, Cukrovarnicka 10, 16200 Prague, Czech Republic*
[3]*Institute for Materials Research, Tohoku University, Sendai, Miyagi 980-8577, Japan*

**Abstract**

Photo- and thermally stimulated luminescence characteristics of $Gd_3(Ga,Al)_5O_{12}$:Ce single crystals co-doped with W and Mo are investigated in the 85 - 510 K temperature range and compared with the corresponding characteristics of the undoped and $Ce^{3+}$ - doped $Gd_3(Ga,Al)_5O_{12}$ single crystals of similar composition. A strong effect of the W and Mo impurity ions appears in the photoluminescence spectra and temperature dependences of the photoluminescence intensity, afterglow intensity and decay kinetics, thermally stimulated luminescence (TSL) intensity and TSL glow curves, excitation spectra of the TSL glow curve peaks and activation energy of their creation. The obtained results are explained by the enhancement of the intrinsic emission contribution into the luminescence spectrum of $Gd_3(Ga,Al)_5O_{12}$:Ce,W and $Gd_3(Ga,Al)_5O_{12}$:Ce,Mo due to a large concentration of various W - and Mo - related electron traps in these crystals and, consequently, the $O^-$ - type hole centers, as well as intrinsic crystal lattice defects (e.g., cation vacancies needed for the excess positive charge compensation of the $W^{6+}$ and $Mo^{6+}$ ions). The influence of the co-doping with W and Mo ions on the scintillation characteristics of $Gd_3(Ga,Al)_5O_{12}$:Ce is discussed.

## 1. Introduction

Scintillating materials are currently widely used for radiation detection in many fields, such as medical imaging, high energy physics calorimetry, bolometry for rare events search, industrial control, safety and homeland security, and others [1,2,3]. Among them, wide bandgap oxide dielectrics of high degree of structural perfection are the most suitable for such a purpose [4]. Most of the applications using scintillating materials require high density material and efficient and fast scintillation mechanism. Based on these requirements, an impressive number of heavy cation (particularly lutetium/yttrium/gadolinium)- based hosts doped with $Ce^{3+}$ or $Pr^{3+}$ were developed. $Lu_3Al_5O_{12}$ (LuAG) host appeared as very promising scintillator with both Ce and Pr doping and excellent energy resolution for the latter. Systematic effort has been devoted to further improve their time characteristics by an admixture of gallium [5,6,7], the effect of which was explained by the removal of shallow electron traps due to the shift of the conduction band edge to lower energies [8]. By composition engineering (tailoring), a new family of materials, so-called multicomponent garnet scintillators, has been recently developed. Balanced admixture of Ga and Gd into YAG and LuAG, with a general chemical formula (Gd,Lu,Y)$_3$(Ga,Al)$_5$O$_{12}$:Ce, so-called GAGG:Ce, resulted in suppressed trapping effect and enormously increased light yield up to 50000-60000 ph/MeV [6,2].

Another approach in improving light yield and timing characteristics of the scintillation crystals is based on their co-doping with optically inactive ions. In particular, in LuAG:Ce, the codoping by Me$^{2+}$ (Me=Mg, Ca) dramatically increased the light yield ([7] and refs. therein). It was explained by increase of Ce$^{4+}$ concentration as the Ce$^{4+}$ states provide effective and fast single scintillation cycle in addition to the conventional Ce$^{3+}$ scintillation. Codoping of GAGG:Ce by divalent Ca and Mg ions always resulted in light yield decrease. However, such co-doping markedly reduces the rise time in its scintillation responce and, thus, improves timing resolution [9,10].

In the case of LuAG:Pr a favorable effect of Mo co-doping has been demonstrated recently [11]. The light yield increased up to 26200 ph/MeV as compared to 17000-19000 ph/MeV in Mo undoped crystals and no significant loss in timing response was observed. This strategy of Mo co-doping was recently applied to GAGG:Ce crystals, namely, W-codoped Gd$_3$Ga$_3$Al$_2$O$_{12}$:Ce crystals were grown by the micro-pulling down method and their scintillation characteristics were tested [12]. Similar to LuAG:Pr,Mo, the light yield increased at W concentrations of 200-500 ppm but decreased at W concentrations above ~1000 ppm. There was again no significant change in the timing characteristics of scintillation. It was proposed that W-ion substitution in the octahedral garnet site can prevent the creation of anti-site defects or reduce electron traps concentration caused by oxygen vacancies. Obviously, this co-doping positive effect should be studied further to obtain detailed explanations of the mechanism responsible for light yield increase and the relationship between W and Mo co-doping and defects behavior.

It is worth also to mention that in the Czochralski growth of aluminum garnets the crucibles made from Mo or W are often used. These ions can penetrate into crystals from the crucible material [13]. The influence of these dopants on scintillation characteristics (creation/suppression of defects) is thus an important task in order to further optimize the growth technology of garnet-based scintillators.

In the crystal lattice of Gd$_3$(Ga,Al)$_5$O$_{12}$ (GAGG), the impurity W and Mo ions substitute, most probably, for the larger Ga$^{3+}$ ions in the octahedral sites [14,15]. It is expected that both these impurities, can partly penetrate into the lattice as Mo$^{3+}$ and W$^{3+}$ ions due to similar to Ga$^{3+}$(Al$^{3+}$) valence. For instance, the Mo$^{3+}$ ions were detected in YAG [16] and LuAG:Sc [17] crystals grown in Mo crucibles. Mo$^{3+}$ and W$^{3+}$ are paramagnetic ions and their charge state can be easy determined from Electron Paramagnetic Resonance (EPR) spectra. But, this is not case of GAGG crystals which contain paramagnetic Gd$^{3+}$ cations. In these crystals, spins of any paramagnetic impurity will be exchange-coupled to Gd$^{3+}$ spins and only a common spectrum will be observed like in a magnetic material [18]. To the best of our knowledge, neither Mo$^{3+/4+/5+}$ nor W$^{3+/4+/5+}$ EPR spectra were observed in YAG/LuAG crystals doped with Ce. Therefore, we assume that majority of W and Mo ions in GAGG lattice are in the 6+ charge state. Ionic radii of both the Mo$^{6+}$ and W$^{6+}$ (0.59 Å and 0.60 Å, respectively) are very close to that of $Ga^{3+}$ (0.62 Å) in the octahedral site [19]. This is also the most stable valence of Mo and W ions.

The aim of this work was to investigate the influence of the W and Mo dopants with a large excess positive charge on the photoluminescence and thermally stimulated luminescence (TSL) characteristics as well as on the processes of the electron and hole centers creation under UV irradiation of the $Ce^{3+}$-doped GAGG crystals. The results obtained for the GAGG:Ce,W single crystals, where the co-doping effect is found to be particularly strong [12], are considered in more detail. The data obtained for the single crystals of GAGG:Ce,Mo are shortly presented as well.

## 2. Experimental procedure

The single crystals of $Gd_3Ga_xAl_{5-x}O_{12}$:1% Ce, 0.1% W (denoted as GAGG:Ce, 0.1% W), $Gd_3Ga_xAl_{5-x}O_{12}$:1% Ce, 0.3% W (GAGG:Ce, 0.3% W), $Gd_3Ga_xAl_{5-x}O_{12}$:1% Ce, 0.1% Mo (GAGG:Ce, 0.1% Mo), $Gd_3Ga_xAl_{5-x}O_{12}$:1% Ce, 0.3% Mo (GAGG:Ce, 0.3% Mo), and $Gd_3Ga_xAl_{5-x}O_{12}$:1% Ce (GAGG:Ce) of similar host composition and the same $Ce^{3+}$ concentration investigated in this work were grown in Japan, Tohoku University, Sendai using the Czochralski technique (details about the growth technology can be found in [20]). The host compositions of some of these crystals, calculated from the data on mass concentrations of Al, Ga, Gd determined by electron microprobe X-ray microanalysis, are presented in Table 1. For this purpose, electron microprobe JEOL model JXA 8230 was used. Pure single crystals $Gd_3Al_5O_{12}$ and $Y_3Al_5O_{12}$ were used as calibration standards.

**Table 1**. Host composition of some of the investigated crystals (oxygen content is fixed at stoichiometric value).

| Crystal | Host composition |
|---|---|
| GAGG:Ce, 0.1% W | $Gd_{2.81-2.85}Ga_{2.95-3.00}Al_{2.18-2.20}O_{12}$ |
| GAGG:Ce, 0.1% Mo | $Gd_{2.82-2.85}Ga_{2.69-2.75}Al_{2.43-2.46}O_{12}$ |
| GAGG:Ce | $Gd_{2.82-2.86}Ga_{2.72-2.74}Al_{2.42-2.44}O_{12}$ |

For comparison, the characteristics of the undoped $Gd_3Ga_xAl_{5-x}O_{12}$ (x = 3) and the $Gd_3Ga_xAl_{5-x}O_{12}$:0.035% Ce (x = 2.83) single crystals, grown by the Czochralski method in Prague, Institute of Physics and studied in our previous works [21,22,23,24], were measured at the same experimental conditions. These crystals are denoted as the GAGG (Prague) and GAGG:Ce (Prague) crystals.

The steady-state emission and excitation photoluminescence (PL) spectra in the 85 - 500 K temperature range were measured using a setup, consisting of the LOT - ORIEL xenon lamp (150 W), two monochromators (SF-4 and SPM-1) and a nitrogen cryostat. The luminescence was detected by a photomultiplier (FEU-39 or FEU-79) connected with an amplifier and recorder. The spectra were corrected for the spectral dependence of the excitation light intensity, transmission and dispersion of the monochromators, and spectral sensitivity of the photodetectors.

Thermally stimulated luminescence (TSL) glow curves $I_{TSL}(T)$ were measured with a heating rate of 0.2 K/s in the 85 - 510 K temperature range after selective irradiation of the crystals at different temperatures $T_{irr}$ with different irradiation photon energies $E_{irr}$ varying in the 2.4 - 4.7 eV energy range. A crystal located in a nitrogen cryostat was irradiated with the LOT - ORIEL xenon lamp (150 W) through a monochromator SF-4. The TSL glow curves were measured with the monochromator SPM-1 and detected with the photomultiplier FEU-39 and recorder. For each TSL glow curve peak, the TSL peak excitation spectrum (also called as the TSL peak creation spectrum), i.e., the dependence of the TSL glow curve peak intensity ($I_{TSL}^{max}$) on the irradiation photon energy $E_{irr}$, was measured. These spectra were corrected for the irradiation intensity. From the slope of the dependence of the TSL glow curve peak intensity ($I_{TSL}^{max}$) on the irradiation temperature $T_{irr}$ presented in the ln ($I_{TSL}^{max}$) - $1/T_{irr}$ coordinates, the activation energy $E_a$ for the TSL peak creation was determined.

To determine the trap depth $E_t$ corresponding to each TSL peak, the partial cleaning method was used (for more details, see, e.g., [25] and references therein). The crystal, irradiated at the temperature $T_{irr}$, was heated up to a temperature $T_{stop}$ with a heating rate of 0.2 K/s, then quickly cooled down to $T_{irr}$, and after that the TSL glow curve was recorded. In the next cycle, the same procedure was repeated for the different temperature $T_{stop}$, etc. From the slope of the ln ($I_{TSL}$) as a function of the reciprocal temperature (1/T), the $E_t$ value was calculated.

### 3. Experimental results obtained at the investigation of the GAGG:Ce,W crystals

*3.1. Photoluminescence characteristics*

Unlike the GAGG:Ce (Prague) crystal investigated in [21-24] and the GAGG:Ce crystal investigated in this work, the position of the $Ce^{3+}$ - related emission band of the GAGG:Ce, 0.1% W crystal depends on the excitation energy $E_{exc}$ (see Fig. 1).

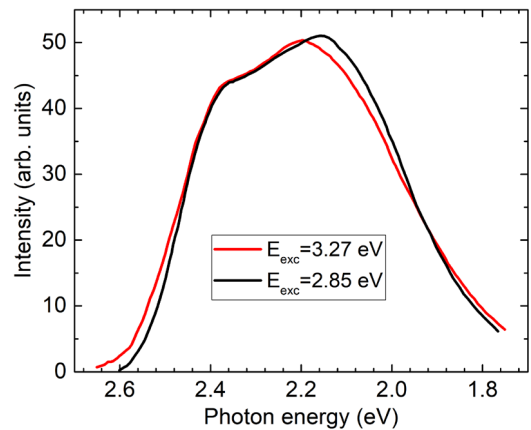

**Fig. 1.** Normalized PL emission spectra of the GAGG:Ce, 0.1% W crystal measured at 85 K under excitation in the $4f – 5d_1$ absorption band of $Ce^{3+}$ ($E_{exc}$ = 2.85 eV) (black line) and in the energy range between the $4f – 5d_2$ and $4f – 5d_1$ absorption bands of $Ce^{3+}$ ($E_{exc}$ = 3.27 eV) (red line).

Under excitation at 85 K in the maximum of the 4f – $5d_1$ absorption band of $Ce^{3+}$ ($E_{exc}$ = 2.85 eV, see Fig. 2a), the doublet emission band is centered at 2.21-2.22 eV (Fig.1, black line). The same spectrum appears under excitation in the $Gd^{3+}$ - related absorption bands (around 4.0 eV and 4.5 eV, Fig. 2a). In the 85 - 295 K temperature range, the emission band position is practically independent of temperature.

However, under excitation at 85 K in the energy range between the 4f – $5d_2$ and 4f – $5d_1$ absorption bands of $Ce^{3+}$ ($E_{exc}$ = 3.27 eV, see Fig. 2a), the emission band is shifted to higher energies with respect to the previous one (see Fig. 1, red line). The same effect is observed in the GAGG:Ce, 0.3% W crystal. The shift is most probably caused by a considerable contribution of the intrinsic electron recombination luminescence into the luminescence spectrum of GAGG:Ce,W measured under this excitation. Indeed, as evident from [21,22], the emission spectra of the undoped and $Ce^{3+}$ - doped GAGG (Prague) crystals are close, but the emission band of the GAGG (Prague) crystal is much broader and located at higher energy (around 2.37 eV [22]) as compared with the emission band of the GAGG:Ce (Prague) crystal. Besides, the intrinsic emission is stimulated with comparable efficiency in a wide energy range, including $E_{exc}$ = 3.27 eV [21]. In the W - free GAGG:Ce crystals, the emission spectra measured under 3.27 eV excitation and under excitation in all the $Ce^{3+}$ - related and $Gd^{3+}$ - related absorption bands coincide. Therefore, the data obtained indicate that the contribution of the intrinsic emission into the luminescence spectrum of GAGG:Ce,W is much larger as compared with GAGG:Ce.

Excitation spectra of the $Ce^{3+}$ emission in the GAGG:Ce, 0.1% W (Fig. 2a) and GAGG:Ce crystals are similar. Besides the most intense $Ce^{3+}$ - related bands located around 3.6 eV and 2.8 eV and arising from the 4f → $5d_2$ and 4f → $5d_1$ transitions of a $Ce^{3+}$ ion, respectively, the spectra contain the narrow $Gd^{3+}$ - related bands due to an effective $Gd^{3+}$ → $Ce^{3+}$ energy transfer.

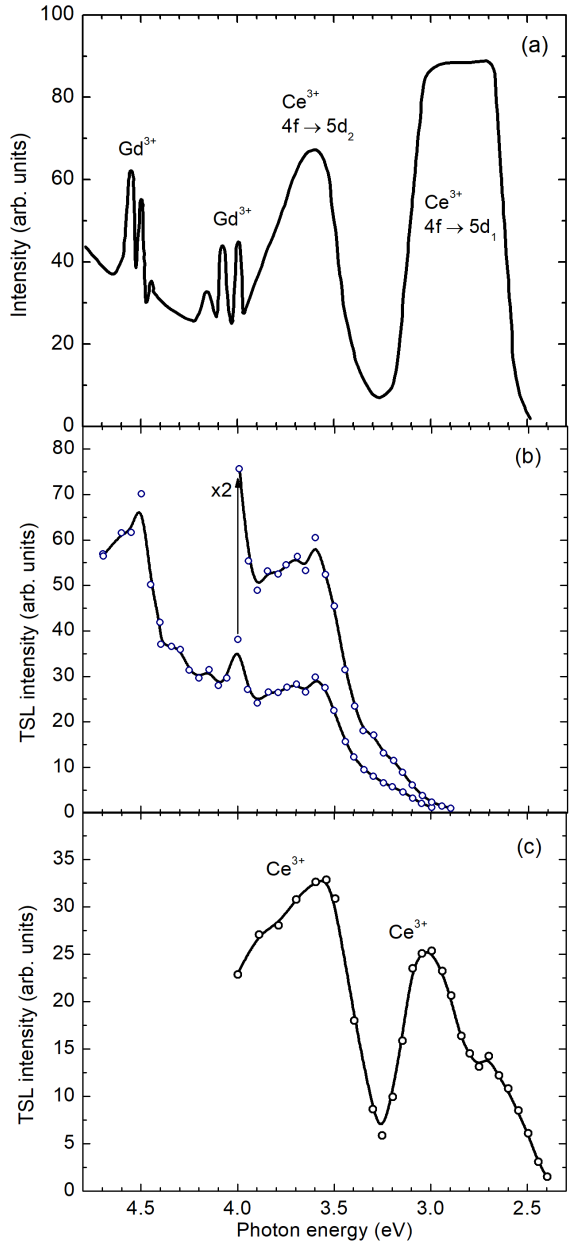

**Fig. 2.** (a) Excitation spectrum of the $Ce^{3+}$ - related emission of the GAGG:Ce, 0.1% W crystal measured at 85 K. (b,c) Dependences of the TSL glow curve peak intensity on the irradiation energy (the TSL peak excitation spectra) measured (b) at 85 K for the 250 K peak and (c) at 295 K for the 370 K peak of the same crystal. $E_{em}$ = 2.3 eV.

As the temperature increases, the intensity (I) of the $Ce^{3+}$ - related emission of the GAGG:Ce, 0.1% W crystal excited around 2.85 eV remains practically constant up to room temperature (RT) and then decreases twice around $T_q$ = 370 K (Fig. 3a, black line). From the temperature dependence of the emission intensity presented in the lnI – 1/T coordinates, the activation energy $E_q$ of this process can be determined. This dependence becomes linear only at T > 420 K, and the corresponding value of $E_q$ is 0.52 ± 0.02 eV (see the inset to Fig. 3a). In the 300 - 400 K temperature range, the luminescence intensity decreases about 5 times with different, much smaller activation energies (see also [22,23]).

Under excitation with $E_{exc}$ = 3.27 eV, the activation energy of the luminescence thermal quenching in GAGG:Ce,W is also much smaller and the $T_q$ value is larger as compared with that obtained under 2.85 eV excitation (see Fig. 3a, red dashed line). The comparison with the I(T) dependence obtained in [22] for the undoped GAGG (Prague) crystal ($T_q$ = 430 K, $E_q$ = 0.11 eV, see also Table 2) indicates that this effect can be explained by a noticeable contribution of the intrinsic emission band into the luminescence spectrum of GAGG:Ce,W. This effect is surely connected with the presence of $W^{6+}$ ions in GAGG:Ce,W as in the $W^{6+}$ - free GAGG:Ce crystal, the I(T) dependences measured under 2.85 eV and 3.27 eV excitations coincide.

The I(T) dependence presented in Fig. 3a (black line) is measured after preliminary heating of the as-received GAGG:Ce, 0.1% W crystal up to 510 K and cooling down to 85 K to destroy the defects optically created during the crystal stay at the daylight at RT. However, we have found that the I(T) dependence (especially the $T_q$ value) of GAGG:Ce,W is strongly sensitive to the pre-history of the sample, namely, to the number and conditions of its preliminary irradiations and heatings up to 510 K. This effect is illustrated in Fig. 3b for the GAGG:Ce, 0.3% W crystal. Indeed, for the I(T) dependence of the as-received crystal (black solid line), $T_q$ = 374 K and $E_q$ = 0.44 eV. After heating of this crystal up to 510 K and cooling down to 85 K, the I(T) dependence is shown by a black dashed line with $T_q$ = 380 K, $E_q$ = 0.47 eV. The I(T) dependence, obtained after eight I(T) measurements at the crystal heating and cooling down in the 410-510 K temperature range carried out for the precise determination of the $E_q$ value, is shown by a red line with $T_q$ = 394 K, $E_q$ = 0.46 eV. After two weeks of various experiments at this crystal, $T_q$ increases up to 406 K (blue line), while the value of $E_q$ remains the same ($E_q$ = 0.46 eV). In should be noted that these results are observed in spite of the fact that all the I(T) measurements were carried out with the narrowest possible monochromator slits and under short irradiations by the excitation light to minimize the photostimulated defects creation during the I(T) measurements. The photoluminescence intensity at 85 K remains practically the same after all these experiments.

Unlike the GAGG:Ce,W crystals, in the GAGG:Ce crystal, the I(T) dependence presented in Fig. 3c is not sensitive to the pre-history of the crystal ($T_q$ = 366 K, $E_q$ = 0.47 ± 0.01 eV).

The $T_q$ and $E_q$ values obtained for different GAGG:Ce,W, GAGG:Ce and GAGG crystals of similar composition are presented in Table 2. Most probably, the $E_q$ energy corresponds to the energy distance ($E_{dc}$) between the relaxed $5d_1$ excited state of $Ce^{3+}$ and the conduction band which should strongly decrease with the increasing Ga content [6-9]. Therefore, the larger $E_q$ value (0.52 eV) in the GAGG:Ce, 0.1% W crystal (x = 2.95 - 3.00) as compared with the $E_q$ value (0.47 eV) in the GAGG:Ce crystal (x = 2.72 - 2.74) cannot be connected with the difference in the Ga content. Probably, it is caused by the perturbation of the energy levels of a $Ce^{3+}$ ion in GAGG:Ce,W by a closely located W ion.

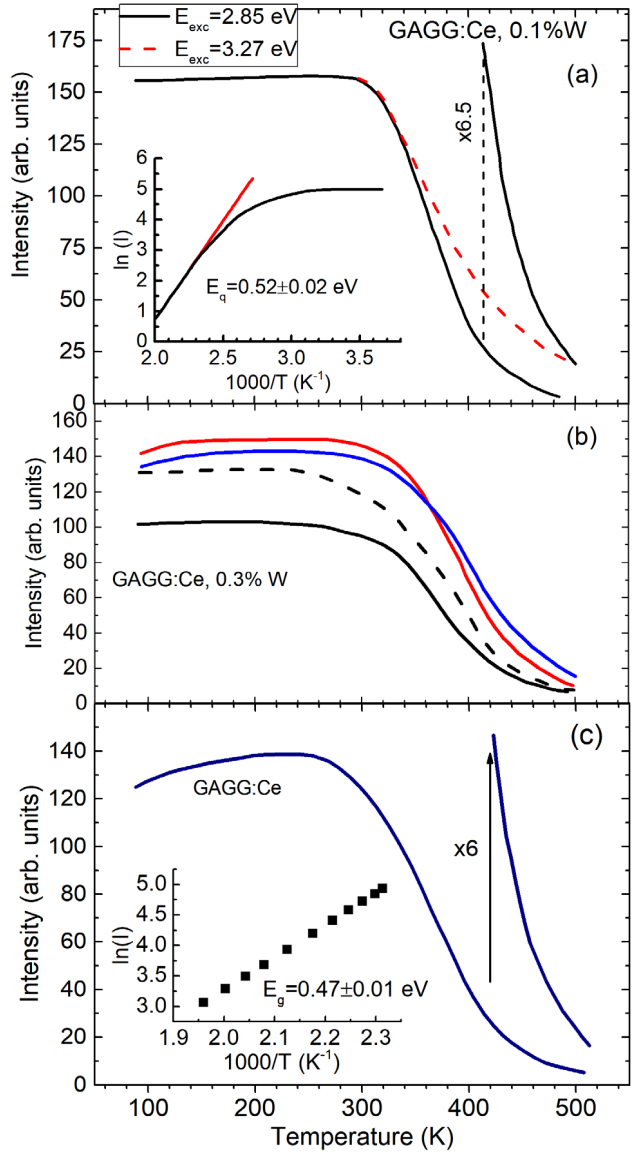

**Fig. 3**. Temperature dependences of the $Ce^{3+}$ - related emission intensity – I(T) measured (a) for the GAGG:Ce, 0.1% W crystal under excitation in the maximum of the $4f – 5d_1$ absorption band of $Ce^{3+}$ ($E_{exc}$ = 2.85 eV, black line) and with $E_{exc}$ = 3.27 eV (red dashed line); (b) under $E_{exc}$ = 2.85 eV for the as-received GAGG:Ce, 0.3% W crystal (black solid line), after heating of this crystal up to 510 K and cooling down to 85 K (black dashed line), after eight I(T) measurements in the 410 - 510 K temperature range (red line), and after two weeks of experiments at this crystal (blue line); (c) for the GAGG:Ce crystal, $E_{exc}$ = 2.85 eV. In the insets to Figs. 3a and 3c, the I(T) dependences measured under 2.85 eV excitation are presented in the ln(I) – 1/T coordinates.

**Table 2.** The comparison of some characteristics of the GAGG:Ce,W, GAGG:Ce,Mo, GAGG:Ce and GAGG single crystals (SC) and the GAGG:Ce epitaxial film (LPE) of similar composition. $T_q$ is the temperature where the emission intensity decreases twice. $E_q$ is the activation energy for the luminescence thermal quenching at T > 420 K. $E_a$ is the activation energy of the TSL glow curve peaks creation under $4f – 5d_1$ excitation.

| Sample | | x | $T_q$, K | $E_q$, eV | $E_a$, eV | Ref. |
|---|---|---|---|---|---|---|
| $Gd_3Ga_xAl_{5-x}O_{12}$:0.035%Ce (Prague) | SC | 2.83 | 360 | 0.48 | 0.21 | this work |
| $Gd_3Ga_xAl_{5-x}O_{12}$ (Prague) | SC | 3.00 | 430 | 0.11 | 0.11 | [2] |
| $Gd_3Ga_xAl_{5-x}O_{12}$:0.15%Ce (Prague) | LPE | 2.70 | 320 | 0.50 | 0.20 | [2] |
| $Gd_3Ga_xAl_{5-x}O_{12}$:1% Ce, 0.1% W | SC | 2.95-3.00 | 368-405 | 0.52 | 0.25 | this work |
| $Gd_3Ga_xAl_{5-x}O_{12}$:1% Ce, 0.3% W | SC | - | 374-407 | 0.46 | 0.28 | this work |
| $Gd_3Ga_xAl_{5-x}O_{12}$:1% Ce, 0.1% Mo | SC | 2.69-2.75 | 360-396 | 0.49 | 0.27 | this work |
| $Gd_3Ga_xAl_{5-x}O_{12}$:1% Ce, 0.3% Mo | SC | - | 360-406 | 0.48 | 0.27 | this work |
| $Gd_3Ga_xAl_{5-x}O_{12}$:1% Ce | SC | 2.72-2.74 | 366 | 0.47 | 0.31 | this work |

*3.2. Afterglow*

After irradiation of the GAGG:Ce, 0.1% W crystal at 85 K with $E_{irr}$ = 3.6 eV, a strong afterglow is observed. The afterglow intensity is much larger and the afterglow decay is much slower (Fig. 4, black solid line) as compared with the W - free GAGG:Ce (Prague) crystal of approximately the same composition (black dashed line). Similar afterglow decay kinetics are observed also for the GAGG:Ce, 0.3% W and GAGG:Ce crystals, respectively. This effect can also be caused by the W-induced considerable contribution of the slow intrinsic electron recombination luminescence into the luminescence spectrum of GAGG:Ce,W. After irradiation at 295 K with $E_{irr}$ = 2.85 eV, much faster afterglow appears as compared with that at 85 K (red line).

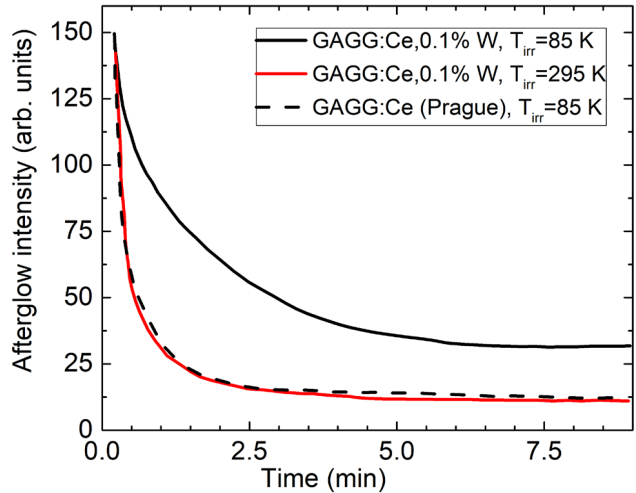

**Fig. 4**. The normalized afterglow decay curves of the GAGG:Ce, 0.1% W crystal irradiated at 85 K (black solid line) and at 295 K (red line) with $E_{irr}$ = 3.6 eV and $E_{irr}$ = 2.85 eV, respectively. For comparison, the afterglow decay curve of the GAGG:Ce (Prague) crystal irradiated at 85 K with $E_{irr}$ = 3.6 eV (black dashed line) is also shown. $E_{em}$ = 2.3 eV.

*3.3. TSL glow curves*

After irradiation at 85 K, several times more intense TSL is observed in the GAGG:Ce,W crystals as compared with the W - free GAGG:Ce crystal. The strongly dominating TSL glow curve peak of GAGG:Ce, 0.1% W is located at 106 K (Fig. 5a). The position of this peak is found to be practically independent of the irradiation energy $E_{irr}$. Much less intense TSL glow curve peaks are located at about 135 K, 185 K, 250 K, 370 K, and 455 K. The approximate intensity ratio of the TSL peaks at 106 K, 185 K, 250 K, 370 K and 455 K is 100 : 5 : 10 : 1 : 0.25 and does not depend on $E_{irr}$.

The TSL glow curves measured at the same conditions for the GAGG:Ce, 0.3% W and GAGG:Ce crystals are presented in Figs. 5b and 5c, respectively. The comparison of Figs. 5a and 5b allows to conclude that the positions of some TSL glow curve peaks (presented in Table 3) depend on the $W^{6+}$ content. Indeed, as the concentration of $W^{6+}$ increases, the most intense TSL peak is shifting from 106 K to 108 K and becomes broader as the relative TSL intensity around 135 K increases. The higher-temperature TSL peaks are also shifting to higher temperatures (e.g., from 370 K to 380 K, from 455 K to 472 K). The intensity ratio of the ≈250 K and ≈185 K TSL peaks increases from 2.15 to 2.6 - 2.7. The intensity ratio of the 370 - 380 K and 455 - 472 K TSL peaks decreases from 4.5 to 2.3 after irradiation at 295 K and from 1.9 to 1.5, after irradiation at 85 K. It seems so that the increasing $W^{6+}$ content results in the increase of relative intensity of the TSL peaks located around ≈ 135 K, ≈ 250 K, and ≈ 470 K.

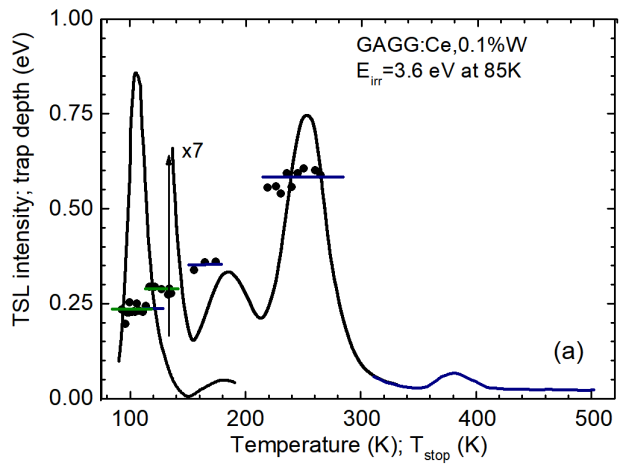
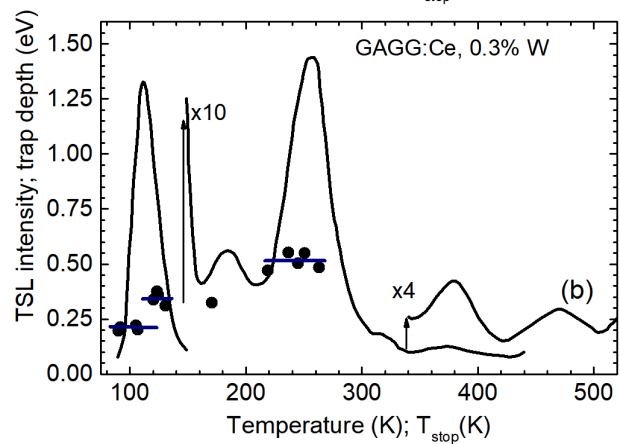
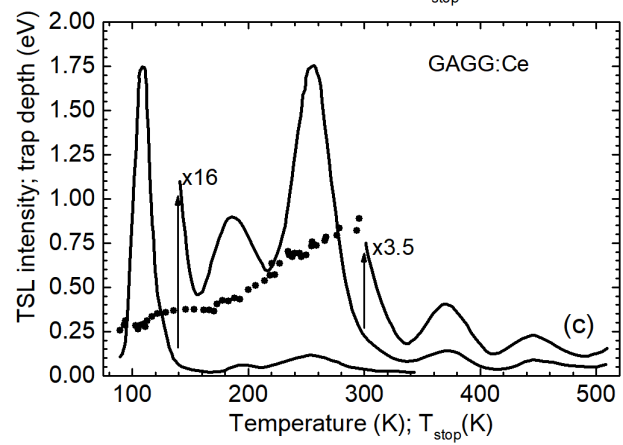
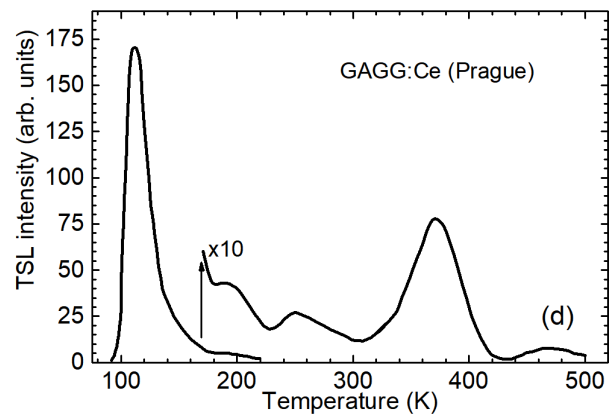

**Fig. 5.** TSL glow curves measured after irradiation of the (a) GAGG:Ce, 0.1% W, (b) GAGG:Ce, 0.3% W, (c) GAGG:Ce, and (d) GAGG:Ce (Prague) crystals at 85 K with $E_{irr}$ = 3.6 eV. The dependences of the trap depths $E_t$ on the temperature $T_{stop}$ measured after irradiation at 85 K are shown by filled circles.

Unlike the W - free samples, a noticeable reduction of the TSL intensity in the course of repeated UV irradiation and heating up to 510 K is found in the GAGG:Ce,W crystals. This effect considerably increases with the increasing W content, especially under irradiation in the energy ranges where the $Ce^{3+}$ absorption is relatively weak (e.g., in the $Gd^{3+}$- related absorption bands). Namely, during two weeks of experiments, the TSL intensity of GAGG:Ce, 0.3% W decreases gradually 2 - 3 times. The intensity of the 252 K peak decreases much more than the intensity of the 186 K peak. As a result, the 252 K/186 K peaks intensity ratio decreases from 2.6 - 2.7 to 2.35. The TSL intensity at T > 500 K increases gradually indicating the enhancement of higher-temperature TSL peaks (not measurable at our setup). However, as mentioned above, no changes appear in the photoluminescence intensity of GAGG:Ce,W measured at 85 K.

It is interesting to note that the positions and especially intensity ratios of the TSL peaks shown in Figs. 5a-5c and Table 3 differ from those obtained at the same conditions for the GAGG:Ce (Prague) single crystal of approximately the same composition and grown by the same method (see Fig. 5d). For example, in the GAGG:Ce (Prague) crystal irradiated at 85 K, the most intense peak is located at 113 K, and this peak is about 1.3 times broader as compared with the 108 K peak in the GAGG:Ce crystal. In the undoped $Gd_3Ga_xAl_{5-x}O_{12}$ crystals also grown by the Czochralski method as well as in the $Gd_3Ga_xAl_{5-x}O_{12}$:Ce LPE films of similar composition (x = 3 and x = 2.7, respectively) prepared in Prague and studied in [21-23], the most intense TSL peak is also located at higher temperature (around 114 - 120 K, see Table 4). The approximate intensity ratio of the 113 K, ≈190 K, 252 K, 372 K, and 470 K peaks in the GAGG:Ce (Prague) crystal is 100 : 2.5 : 1.5 : 4 : 0.4 (see Fig. 5d), while the intensity ratio of the corresponding TSL peaks in the GAGG:Ce crystal is 100 : 3 : 6 : 0.45 : 0.2 (Fig. 5c). The reason of the difference in the TSL characteristics of the Prague and Japan crystals is not clear.

After irradiation of the GAGG:Ce,W crystals at 295 K, the main TSL peaks appear at about 315 K, 372 - 380 K and 455 - 470 K (Fig. 6, black solid lines). The positions and relative intensities of the peaks strongly depend on the W content. Namely, as the W content increases, the intensities of the ≈315 K and 455 - 470 K peaks increase with respect to the 372 - 380 K peak dominating also in the GAGG:Ce crystal (compare Figs. 6a and 6b). In Fig. 6a, a strong difference between the TSL glow curves of the GAGG:Ce (red solid line) and GAGG:Ce (Prague) (red dashed line) crystals measured after irradiation at 295 K is demonstrated.

The dependences of the trap depth $E_t$ values on the temperature $T_{stop}$ obtained after irradiation of the crystals at different temperatures $T_{irr}$ are shown in Figs. 5 and 6 (circles). In assumption on the first-order recombination kinetics, the approximate values of the frequency factors $f_0$ are determined as well.

The positions of the main TSL glow curve peaks ($T_m$) and the corresponding values of $E_t$ and $f_0$ are presented in Table 3.

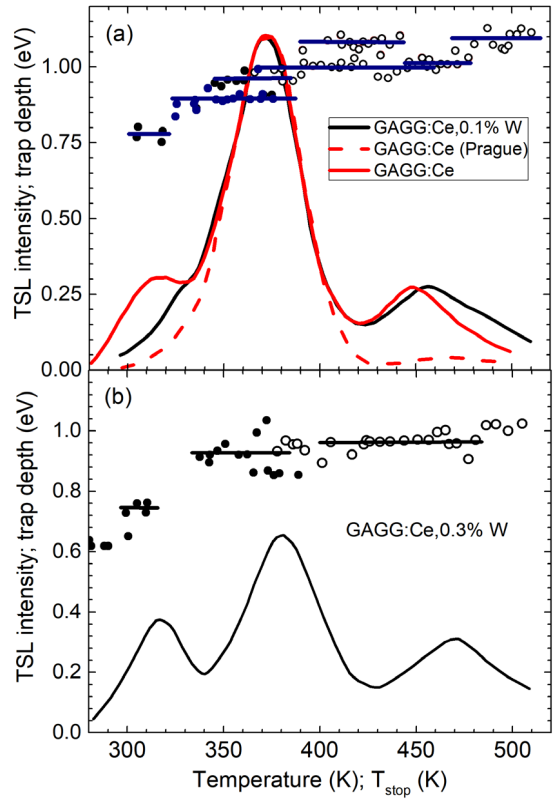

**Fig. 6.** Normalized TSL glow curves measured after irradiation with $E_{irr}$ = 2.85 eV at 295 K (a) for the GAGG:Ce, 0.1% W (black line), GAGG:Ce (red solid line), and GAGG:Ce (Prague) (red dashed line) crystals and (b) for the GAGG:Ce, 0.3% W crystal. Dependences of the trap depths $E_t$ on the temperature $T_{stop}$ measured for these crystals are shown by filled circles ($T_{irr}$ = 295 K) and empty circles ($T_{irr}$ = 380 - 390 K).

**Table 3.** The comparison of the TSL glow curve peaks positions ($T_m$), trap depths ($E_t$), and frequency factors ($f_0$) obtained for the investigated crystals.

| Sample | $T_m$, K | $E_t$, eV | $f_0$, s$^{-1}$ |
|---|---|---|---|
| GAGG:Ce, 0.1% W | 106 | 0.25 | $10^{10}$ |
|  | 135 | 0.32 | $10^{10}$ |
|  | 185 | 0.39 | $10^{9}$ |
|  | 250 | 0.64 | $10^{11}$ |
|  | 370 | 0.95 | $10^{11}$ |
|  | 455 | 1.00 | $10^{9}$ |
| GAGG:Ce, 0.3% W | 108 | 0.25 | $5\times10^{10}$ |
|  | 135 | 0.35 | $5\times10^{11}$ |
|  | 186 | 0.33 | $10^{7}$ |
|  | 252 | 0.62 | $10^{11}$ |
|  | 315 | 0.75 | $10^{10}$ |

|  | 380 | 0.93 | $10^{10}$ |
|---|---|---|---|
|  | 472 | 0.95 | $10^{8}$ |
| GAGG:Ce, 0.1% Mo | 106 | 0.25 | $10^{10}$ |
|  | 130 | 0.30 | $5\times10^{10}$ |
|  | 190 | 0.36 | $10^{8}$ |
|  | 254 | 0.62 | $5\times10^{10}$ |
|  | 315 | 0.68 | $10^{9}$ |
|  | 378 | 0.95 | $10^{11}$ |
|  | 465 | 0.90 | $10^{8}$ |
| GAGG:Ce, 0.3% Mo | 106 | 0.25 | $10^{10}$ |
|  | 130 | 0.35 | $5\times10^{11}$ |
|  | 189 | 0.38 | $10^{8}$ |
|  | 255 | 0.62 | $5\times10^{10}$ |
|  | 310 | 0.70 | $5\times10^{9}$ |
|  | 373 | 0.98 | $10^{11}$ |
|  | 450 | 0.93 | $10^{8}$ |
| GAGG:Ce | 108 | 0.26 | $10^{11}$ |
|  | 130 | 0.35 | $5\times10^{11}$ |
|  | 186 | 0.38 | $5\times10^{8}$ |
|  | 252 | 0.64 | $10^{11}$ |
|  | 315 | 0.78 | $5\times10^{10}$ |
|  | 370 | 0.95 | $10^{11}$ |
|  | 444 | 1.05 | $10^{9}$ |

For GAGG:Ce (Prague) single crystals and epitaxial films of similar composition, these values have been reported in [22,23]. For the dominating complex 113 - 114 K TSL glow curve peak, the values of $E_t \approx 0.20$ eV and $f_0 \sim 10^7 - 10^8$ s$^{-1}$ were obtained. However, in all the crystals investigated in this work, the frequency factor of the most intense 106-108 K peak is found to be much larger ($f_0 \sim 10^{10} - 10^{11}$ s$^{-1}$). This can indicate the increased contribution of an electron origin into the complex TSL peak in this temperature range. This peak can be connected with the presence of W$^{6+}$ ions which are very effective traps for electrons due to their large excess positive charge and the existence of the stable W$^{3+}$ centers. Probably, the TSL glow curve peak at 106 - 108 K in GAGG:Ce,W arises (at least partly) from the electron W-related centers.

Due to a large concentration of W - related electron traps, the intrinsic hole centers should also be effectively created in the irradiated GAGG:Ce,W crystals. Indeed, as the electrons recombination with the intrinsic hole centers results in the appearance of the intrinsic luminescence of GAGG, this fact could explain a large contribution of the intrinsic emission band into the luminescence spectrum of GAGG:Ce,W. According to [24], the TSL peaks of a hole origin should contribute into the complex TSL glow peaks located around 70 K and 220-250 K and observed in the undoped GAGG crystal X-ray irradiated at 9 K. This conclusion has been confirmed by a relatively small values of the frequency factors obtained for both these peaks in the undoped GAGG crystal ($f_0 \sim 10^7$ s$^{-1}$ [21]). However, in the investigated crystals not only the 106 - 108 K peak but also the peak at 250 - 252 K (with relatively large $f_0 \sim 5 \times 10^{10} - 10^{11}$ s$^{-1}$, see Table 3) should be mainly of an electron origin. Only the complex TSL glow curve peak in the 450 - 470 K temperature range, characterized by the $E_t$ value of about 0.9 - 1.0 eV and relatively small value of $f_0 \sim 10^8 - 10^9$ s$^{-1}$, can surely be ascribed to hole centers. Besides, it is also not excluded that the TSL peak at about 185 - 190 K with a relatively small value of $f_0 \sim 10^8 - 10^9$ s$^{-1}$ can arise from hole centers. It should be noted that the origin of the dominating hole centers in GAGG and GAGG:Ce,W can be different. Due to a large excess positive charge of W at different valence, the GAGG:Ce,W crystals should contain much larger number of cation vacancies which can act as efficient traps of holes. These hole centers should be more thermally stable as compared to the self-trapped holes.

### *3.4. Excitation spectra of the TSL glow curve peaks*

In Fig. 2, the dependences of the TSL glow curve peak intensity on the irradiation energy $E_{irr}$ (the TSL peak excitation spectra or the TSL peak creation spectra) presented for the GAGG:Ce, 0.1% W crystal (Figs. 2b,c) are compared with the excitation spectrum of the Ce$^{3+}$ - related emission (Fig. 2a). The spectra are corrected for the irradiation intensity.

At 85 K (Fig. 2b) the creation spectrum is measured for the 250 K peak. It is evident that the TSL peaks appear after irradiation of GAGG:Ce,W not only in the 4f – 5d$_2$ (around 3.6 eV) and higher-energy absorption bands of Ce$^{3+}$ but also in the Gd$^{3+}$ - related absorption bands. The same result was observed also for the undoped GAGG (Prague) crystal and for the GAGG:Ce (Prague) crystal (see Refs. [21,22]). It is interesting to note that unlike the GAGG:Ce (Prague) crystal, the defects in GAGG:Ce,W are created also in the energy range between the 4f – 5d$_2$ and 4f – 5d$_1$ absorption band of Ce$^{3+}$ (around 3.27 eV, see Fig. 2b). At 295 K, the creation spectrum is measured for the 370 K peak (Fig. 2c). At these temperatures, the defects responsible for the TSL glow curve peaks are optically created also in the 4f - 5d$_1$ absorption region of Ce$^{3+}$ centers.

In Fig. 7, the TSL glow peaks excitation spectra measured at 85 K are compared for the GAGG:Ce, 0.1% W (a) and GAGG:Ce (b) crystals as well as for the undoped GAGG (Prague) and GAGG:Ce (Prague) crystals (c). Unlike the GAGG:Ce (Prague) crystal (Fig. 7c, empty circles), the TSL peaks in GAGG:Ce,W are created under irradiation in the energy range between the 4f - 5d$_2$ and 4f - 5d$_1$ absorption bands of Ce$^{3+}$ and even in the range of the 4f - 5d$_1$ band (see, e.g., Fig. 7a; compare also Figs. 2a and 2b) despite the fact that the 5d$_1$ excited level of Ce$^{3+}$ is located well below the bottom of the conduction band of GAGG [22]. As the content of W$^{6+}$ increases, relative TSL intensity in the 2.4 - 3.3 eV energy range increases with respect to that at $E_{irr}$ = 3.6 eV. In the GAGG:Ce crystal, the defects are also created in the 2.4 - 3.3 eV energy range but with smaller efficiency (Fig. 7b).

In the GAGG:Ce,W crystals, the relative intensity of the Gd$^{3+}$ - related bands in the TSL excitation spectrum (with respect to the Ce$^{3+}$ - related 3.6 eV band) increases with the increasing W content. Besides, the TSL glow curve peaks are much more effectively created in the energy range between the $Ce^{3+}$ and $Gd^{3+}$ absorption bands (Fig. 7a) as compared with the GAGG:Ce (Prague) and GAGG (Prague) crystals (Fig. 7c).

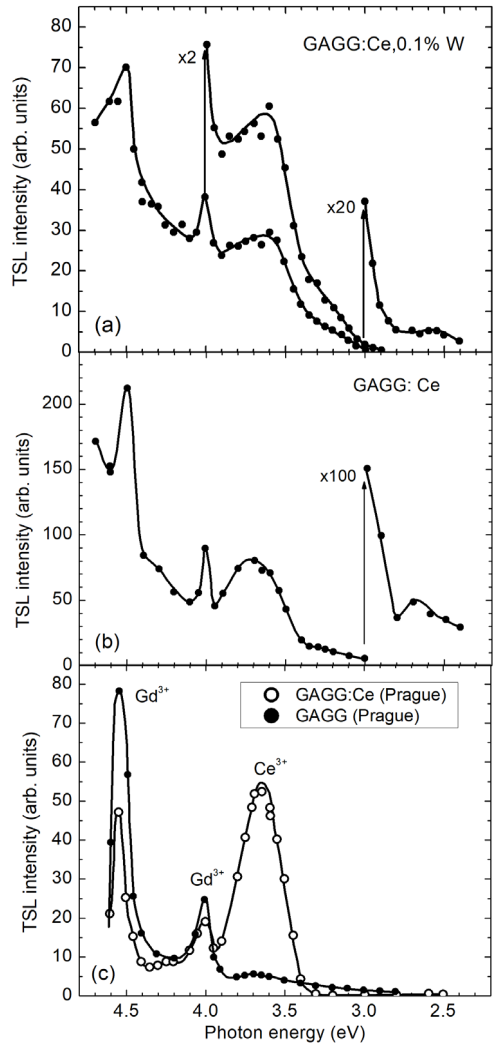

**Fig. 7**. Dependences of the TSL glow curve peak intensity on the irradiation energy $E_{irr}$ measured at 85 K for the 250 K peak in the single crystals of (a) GAGG:Ce, 0.1% W and (b) GAGG:Ce, as well as (c) in the undoped GAGG (Prague) (filled circles) and GAGG:Ce (Prague) (empty circles) crystals. $E_{em}$ = 2.3 eV.

The appearance of the TSL glow curve peaks under irradiation at 85 K in the energy range $E_{irr}$ < 3.4 eV was recently observed in the undoped GAGG (Prague) crystals [21], as well as in GAGG:Ce,Mg single crystals and epitaxial films with a large $Mg^{2+}$ content, where the concentration of $Ce^{3+}$ was considerably reduced due to the $Mg^{2+}$ - induced $Ce^{3+} \rightarrow Ce^{4+}$ transformation [22]. This effect was explained in [21] by the photostimulated electron transfer from the valence band to electron trap levels followed by the creation of intrinsic electron and hole centers and their subsequent recombination accompanied with the intrinsic 2.37 eV emission of GAGG. We suggest that the contribution of the processes characteristic for the undoped GAGG could be considerable in GAGG:Ce,W due to a huge amount of $W^{6+}$ - induced electron traps. Therefore, under irradiation in the energy range where the absorption of $Ce^{3+}$ is sufficiently small (for example, around 3.27 eV and 4.3 eV), the TSL glow curve peaks appear in GAGG:Ce,W mainly due to the processes characteristic for the GAGG host. Indeed, the comparison of Figs. 7a, 7b and 7c indicates that the TSL glow peaks creation spectrum of GAGG:Ce,W (Fig. 7a) looks like an intermediate case between the spectra of the GAGG:Ce (Prague) and undoped GAGG (Prague) crystals (see Fig. 7c). Much smaller relative intensity of the $Ce^{3+}$ - related $4f – 5d_2$ band with respect to the $Gd^{3+}$ - related bands observed in the spectra shown in Figs. 7a and 7b as compared to the spectrum of GAGG:Ce (Prague) shown in Fig. 7c also indicates much larger concentration of intrinsic defects in both the GAGG:Ce,W and the GAGG:Ce crystals. Note that such $Ce^{3+}/Gd^{3+}$ intensity ratio is observed in spite of much large concentration of $Ce^{3+}$ in the latter crystals (see Table 2).

### 3.5. Activation energy of the TSL glow curve peaks creation

The dependences of intensity of the TSL glow curve peak located at 370-380 K ($I_{TSL}$) on the irradiation temperature ($T_{irr}$) presented in the $\ln(I_{TSL}) – 1/T_{irr}$ coordinates measured for the GAGG:Ce,W (filled circles) and GAGG:Ce (empty circles) crystals are shown in Fig. 8. From these dependences, the activation energies $E_a$ of the TSL peak creation under irradiation in the $4f – 5d_1$ absorption band of $Ce^{3+}$ are determined and presented in Table 2.

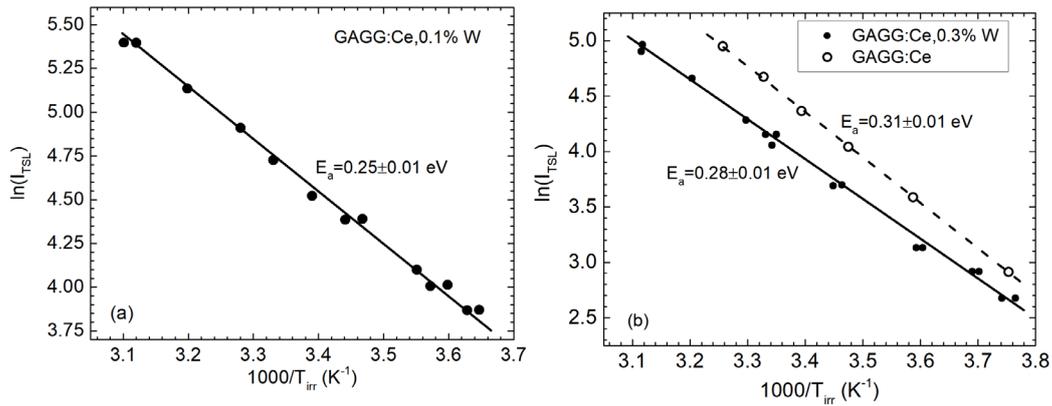

**Fig. 8.** Dependences of the intensity of the TSL glow curve peak located at 370 - 380 K ($I_{TSL}$) on the irradiation temperature ($T_{irr}$) presented in the $\ln(I_{TSL}) – 1/T_{irr}$ coordinates measured (a) for the GAGG:Ce, 0.1% W crystal and (b) for the GAGG:Ce, 0.3% W (filled circles) and GAGG:Ce (empty circles) crystals. $E_{irr}$ = 2.85 eV, $E_{em}$ = 2.3 eV.

It should be noted that the $E_a$ values have been found to be very sensitive to the $Ga^{3+}$ content (x) in $Gd_3Ga_xAl_{5-x}O_{12}$:Ce crystals [26]. Namely, the smaller parameter x is, the larger is the $E_a$ value. Therefore, the difference between the $E_a$ values obtained for the GAGG:Ce (0.31 eV) and GAGG:Ce, 0.1% W (0.25 eV) crystals can be caused by the difference in the Ga content in these crystals (x = 2.72 - 2.74 and x = 2.95 - 3.00, respectively, see Table 2). The same reason can explain the difference between the $E_a$ values obtained for GAGG:Ce, 0.1% W (0.25 eV) and GAGG:Ce, 0.3% W (0.28 eV). Indeed, as the $W^{6+}$ ions substitute most probably for the $Ga^{3+}$ ions in the crystal lattice of GAGG, the Ga content in the crystal with the larger W content could be smaller and, consequently, the $E_a$ value should be larger. Thus, it remains unclear if the co-doping of the GAGG:Ce crystals with $W^{6+}$ influences noticeably the $E_a$ value.

### 4. Experimental results obtained at the investigation of the GAGG:Ce,Mo crystals

#### *4.1. Photoluminescence characteristics*

Similar to the GAGG:Ce,W crystals considered above, a noticeable difference is observed in the photoluminescence spectra of the GAGG:Ce,Mo crystals measured under excitation in the $Ce^{3+}$ - related absorption bands (e.g., with $E_{exc}$ = 2.8 eV, see Fig. 9, black line) and in the energy range between the 4f – $5d_2$ and 4f – $5d_1$ absorption bands of $Ce^{3+}$ (with $E_{exc}$ = 3.27 eV, red line). This effect is surely connected with the presence in the crystals of $Mo^{6+}$ ions as no dependence of the emission spectra on the excitation energy is observed in Mo - free GAGG:Ce crystals (see, e.g., [21-24]).

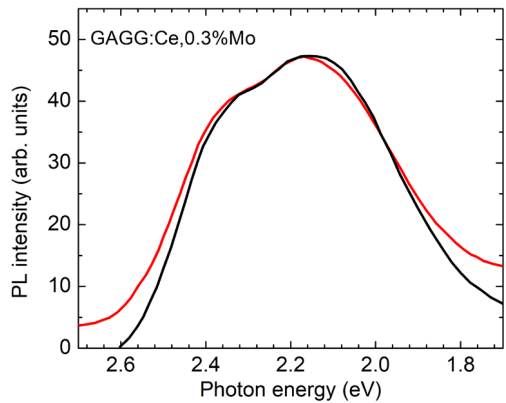

**Fig. 9.** Corrected emission spectra (normalized) of GAGG:Ce, 0.3% Mo measured at 85 K under excitation in the 4f - $5d_1$ absorption band of $Ce^{3+}$ centers ($E_{exc}$ = 2.8 eV) (black line) and in the energy range between the 4f - $5d_2$ and 4f - $5d_1$ absorption bands of $Ce^{3+}$ ($E_{exc}$ = 3.27 eV) (red line).

The excitation spectrum of the $Ce^{3+}$ - related emission of the GAGG:Ce,Mo crystal (Fig. 10) is similar to that obtained for the GAGG:Ce,W and GAGG:Ce crystals.

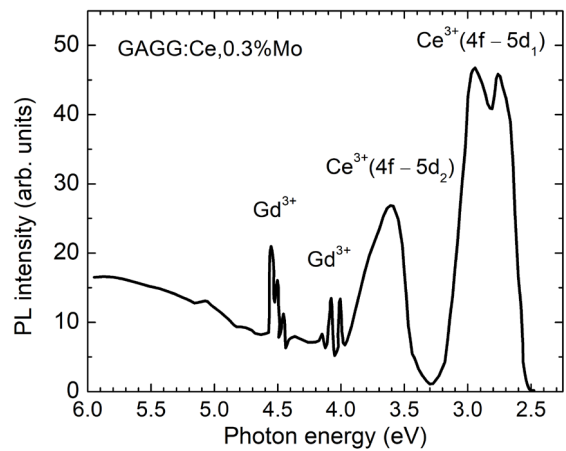

**Fig. 10.** Corrected PL excitation spectrum of GAGG:Ce, 0.3% Mo measured at 85 K for the $Ce^{3+}$- related emission ($E_{em}$ = 2.3 eV).

Similar to the GAGG:Ce,W crystals, temperature dependences of the photoluminescence intensity I(T) of the GAGG:Ce,Mo crystals are sensitive to the pre-history of the investigated sample and the excitation energy (see Fig. 11). The strongest difference in the I(T) dependences is observed in the 300 - 400 K temperature range where this dependence is mainly determined by the defect structure of the crystal. Due to that the temperatures $T_q$, where the photoluminescence intensity decreases twice, differ strongly in the as-received crystals ($T_q$ = 360 K), in the GAGG:Ce, 0.1% Mo crystal heated up to 510 K and cooled down to 85 K ($T_q$ = 386 K), in the same crystal but heated and cooled down many times in the 410 - 510 K temperature range ($T_q$ = 396 K), and in the GAGG:Ce, 0.3% Mo crystal after a month of different experiments ($T_q$ = 406 K). Different $T_q$ values are observed for the same sample under 2.85 eV excitation ($T_q$ = 396 K) and under 3.27 eV excitation ($T_q$ = 426 K). However, no dependence on the pre-history is found for GAGG:Ce (Fig. 11b, black line, $T_q$ = 366 K).

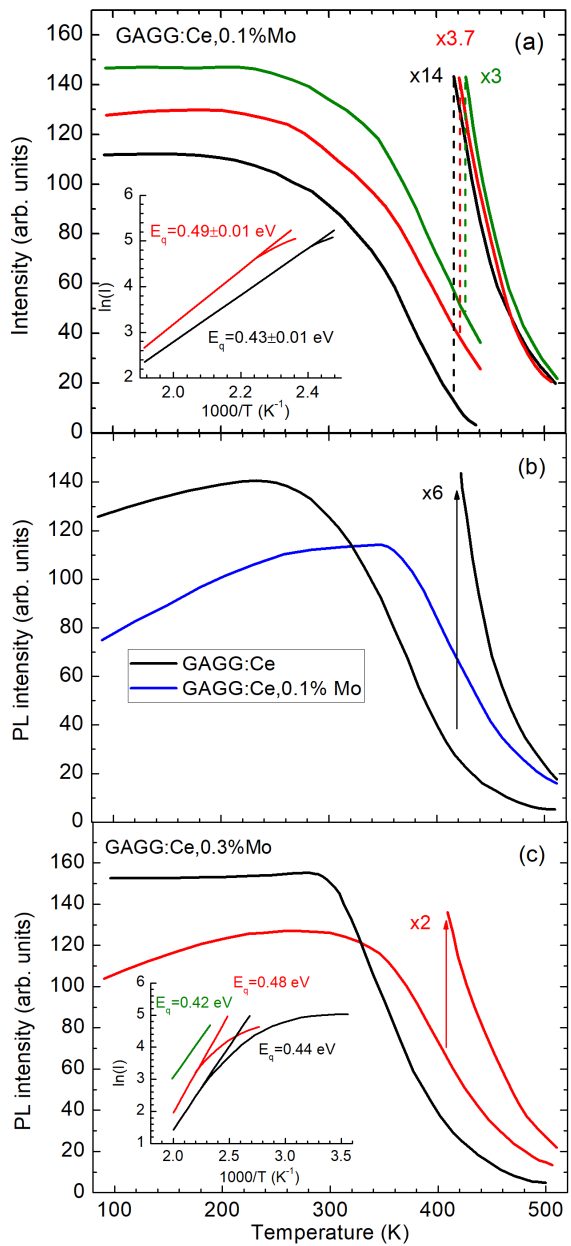

**Fig. 11.** Temperature dependences of the photoluminescence intensity I(T) measured for $E_{em}$ = 2.3 eV (a) under excitation with $E_{exc}$ = 2.85 eV for the as-received GAGG:Ce, 0.1% Mo crystal (black line), after heating this crystal up to 510 K and cooling down to 85 K (red line) and after nine heating/cooling down cycles in the 400 - 500 K temperature range made for the $E_q$ determination (green line). In the inset to Fig. 11a, the temperature dependences measured in the 410 - 510 K temperature range are presented in the ln(I) - 1/T coordinates for the as-received crystal (black line) and after heating the crystal up to 510 K (red line). (b) The I(T) dependence measured for the same crystal with $E_{exc}$ = 3.27 eV after all the I(T) dependences presented in Fig. 11a (blue line) and for the GAGG:Ce crystal with $E_{exc}$ = 2.85 eV (black line). (c) The I(T) dependences measured for the as-received GAGG:Ce, 0.3% Mo crystal (black line) and after one month of experiments with this crystal (red line). In the inset to Fig. 11c, the temperature dependences measured in the 410 - 510 K temperature range are presented in the lnI -1/T coordinates for the as-received crystal (black line), after heating the crystal up to 510 K (red line) and after one month of experiments with this crystal (green line). $E_{exc}$ = 2.85 eV.

From the temperature dependence of the emission intensity measured in the 410 - 510 K temperature range and presented in the ln(I) – 1/T coordinates (see the insets to Figs. 11a and 11c), the activation energy $E_q$ of this process are determined. For the as-grown GAGG:Ce,Mo crystals, the $E_q$ values are usually smaller (0.43 - 0.44 eV) as compared with the samples preliminarily heated up to 510 K and cooled down to 85 K (0.48 - 0.49 eV). The repeated heating cycles up to 410 - 510 K temperature range do not influence considerably the $E_q$ values. This allows us to conclude that these $E_q$ values indicate the energy distance between the $5d_1$ excited level of $Ce^{3+}$ and the conduction band bottom. However, after about a month of experiments, the $E_q$ value noticeably decreases (down to 0.42 eV, see the inset to Fig. 11c, green line). The $T_q$ and $E_q$ values characteristic for the investigated crystals are presented in Table 2.

### 4.2. Afterglow

Similar to the GAGG:Ce,W crystals, the afterglow of GAGG:Ce,Mo is more intense as compared with GAGG:Ce. After irradiation at 85 K in the $Ce^{3+}$ - related absorption bands, the afterglow decay is faster and the contribution of the slow decay components is much smaller (Fig. 12, solid black line) as compared with the irradiation in the energy range (e.g., with $E_{exc}$ = 3.1 eV) where mainly the intrinsic defects can be optically created (dashed black line). In GAGG:Ce,Mo, the afterglow decay is slower and the contribution of the slow decay components is larger (Fig. 12, red line) as compared with GAGG:Ce (green line).

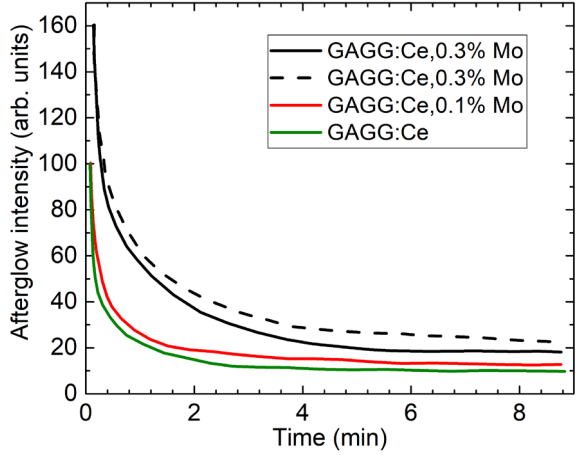

**Fig. 12.** Dependence of the afterglow intensity on the time after irradiation at 85 K with $E_{irr}$ = 3.6 eV (solid lines) and $E_{irr}$ = 3.1 eV (dashed line) measured for the GAGG:Ce, 0.3% Mo crystal (black lines, normalized) as well as for the GAGG:Ce, 0.1% Mo (red line) and GAGG:Ce (green line) crystals (normalized). $E_{em}$ = 2.3 eV.

By analogy with the GAGG:Ce,W crystals, the peculiarities of the photoluminescence and afterglow characteristics of the GAGG:Ce,Mo crystals can be explained by an increased contribution of the intrinsic emission band into their luminescence spectrum due to a large concentration of recombining intrinsic electron and hole centers owing to a large effective positive charge of $Mo^{6+}$ in the crystal lattice of GAGG.

### 4.3. TSL glow curves

The TSL glow curves measured for all the investigated crystals after irradiation at 85 K (Figs. 13a-13c) and 295 K (Fig. 13d) are similar, but the positions of the TSL glow curve peaks (Table 3) and their intensity ratios (Table 4) depend on the crystal. Especially large variations are observed for the two highest-temperature peaks. The intensity ratios of some TSL peaks of GAGG:Ce,Mo are changing also during the measurements (e.g., after one week of experiments, see Table 4). For GAGG:Ce crystals, no such changes are observed. These data also indicate a large concentration of intrinsic defects in the GAGG:Ce,Mo crystals and its dependence on the crystal pre-history.

The dependences of the trap depth $E_t$ values on the temperature $T_{stop}$ obtained after irradiation of a crystal at different temperatures $T_{irr}$ are shown in Fig. 13 (points). In assumption on the first-order recombination kinetics, the approximate values of the frequency factors $f_0$ are determined. The positions of the main TSL glow curve peaks ($T_m$) and the corresponding values of $E_t$ and $f_0$ are presented in Table 3. It is evident that these values are similar to those obtained for the W - co-doped GAGG:Ce crystals and can be interpreted in the same way.

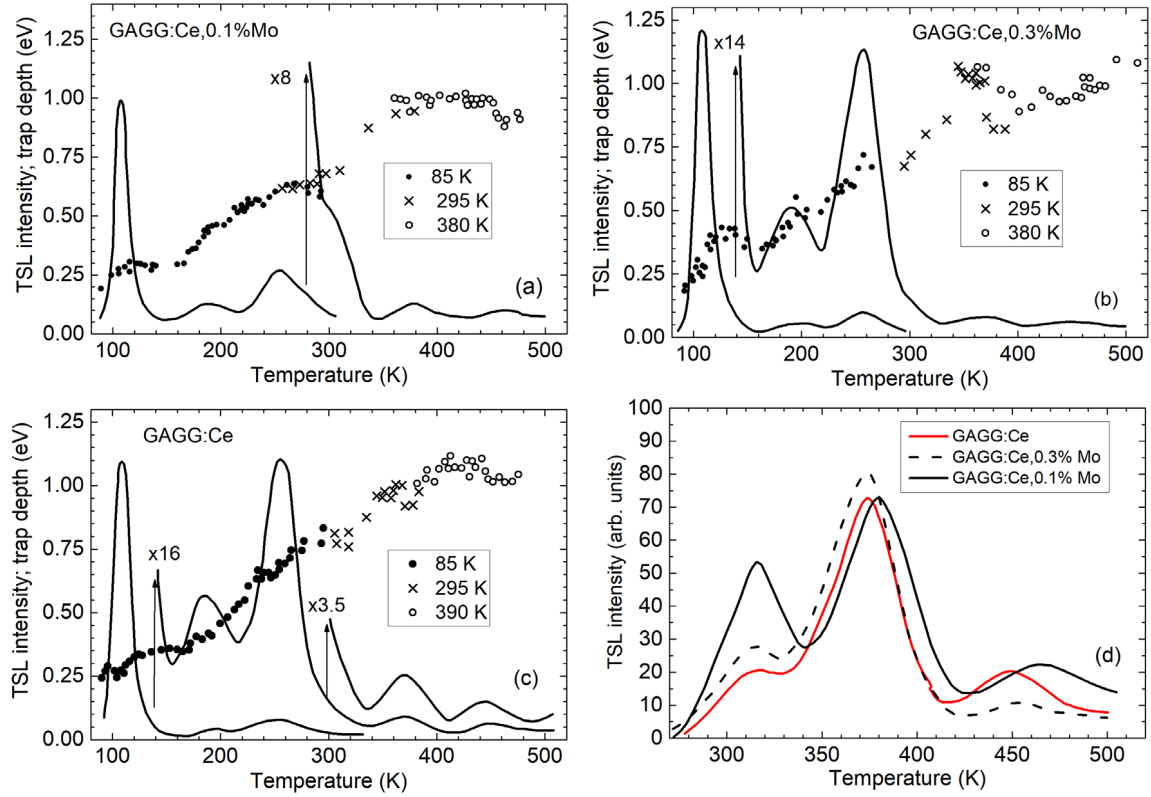

**Fig. 13.** TSL glow curves (solid lines) measured after irradiation of the (a) GAGG:Ce, 0.1% Mo, (b) GAGG:Ce, 0.3% Mo and (c) GAGG:Ce crystals at 85 K in the 4f - $5d_2$ absorption band of $Ce^{3+}$ centers ($E_{irr}$ = 3.6 eV). The dependences of the trap depths $E_t$ on the temperature $T_{stop}$ (red points) measured after irradiation with $E_{irr}$ = 3.6 eV at 85 K (filled circles) and with $E_{irr}$ = 2.65 eV at higher temperatures $T_{irr}$ shown in the legends. (d) TSL glow curves measured after irradiation of the GAGG:Ce, 0.1% Mo crystal (black solid line), GAGG:Ce, 0.3% Mo crystal (black dashed line) and GAGG:Ce crystals (red line) at 295 K in the 4f - $5d_1$ absorption band of $Ce^{3+}$ centers ($E_{irr}$ = 2.65 eV).

**Table 4.** Intensity ratios of the TSL glow curve peaks in the GAGG:Ce, 0.1% Mo, GAGG:Ce, 0.3% Mo, and GAGG:Ce crystals

| Crystal | 376 K/ 470 K | 376 K/ 315 K | 107 K/ 190 K | 107 K/ 255 K | 255 K/ 190 K | 255 K/ 376 K | 376 K/ 468 K |
|---|---|---|---|---|---|---|---|
| 0.1% Mo | 3.7-4.2 | 1.35 | 7-8 | 3.5-3.8 | 2.0-2.3 | 30 | 1.5 |
| after 1 week |  |  | 15 | 7.0-7.2 | 2.4 | 15 |  |
| 0.3% Mo | 10 | 2.9 | 23 | 12 | 2.0-2.4 | 10 | 1.5-2.0 |
| after 1 week | 11 | 2.5-3.0 | 13-14 | 6-9 | 2.5 | 16 |  |
| Mo free | 4.1 | 3.6 | 31 | 16.5 | 1.6 | 14 | 1.9 |

*4.4. Excitation spectra of TSL glow curve peaks*

Excitation spectra of TSL glow curve peaks (defects creation spectra) measured at 85 K (Fig. 14a) and 295 K (Fig. 14b) for the TSL peaks located around 250 - 255 K and 370 - 380 K, respectively, are found to be similar in the investigated GAGG:Ce,Mo (black circles) and GAGG:Ce (red circles) crystals. The spectra are normalized in the $Ce^{3+}$ - related bands (around 3.7 eV, Fig. 14a and around 2.95 eV, Fig. 14b). It is evident that despite much larger $Ce^{3+}$ content in the investigated crystals, relative intensity of the $Gd^{3+}$ - related bands (around 4.5 eV and 4.0 eV) with respect to the $Ce^{3+}$ - related bands in these crystals is much larger as compared with GAGG:Ce (Prague) crystals investigated in [21,22]. Probably, this is caused by larger concentration of intrinsic defects in the investigated crystals, including also the Mo - free GAGG:Ce crystal. Like in the GAGG:Ce,W crystals, the TSL glow curve peaks are created at 85 K also after irradiation in the energy region $E_{exc}$ < 3.3 eV, i.e. in the absorption region where the intrinsic luminescence can be excited (see, e.g., [21]). For the GAGG:Ce,Mo crystals, the efficiency of this process in noticeably larger (Fig. 14a, black circles) as compared with the GAGG:Ce crystal (red circles).

Like in the GAGG:Ce,W crystals, the TSL intensity gradually decreases during the defect creation spectrum measurement. However, in the GAGG:Ce,Mo crystals, the decrease is smaller as compared with GAGG:Ce,W.

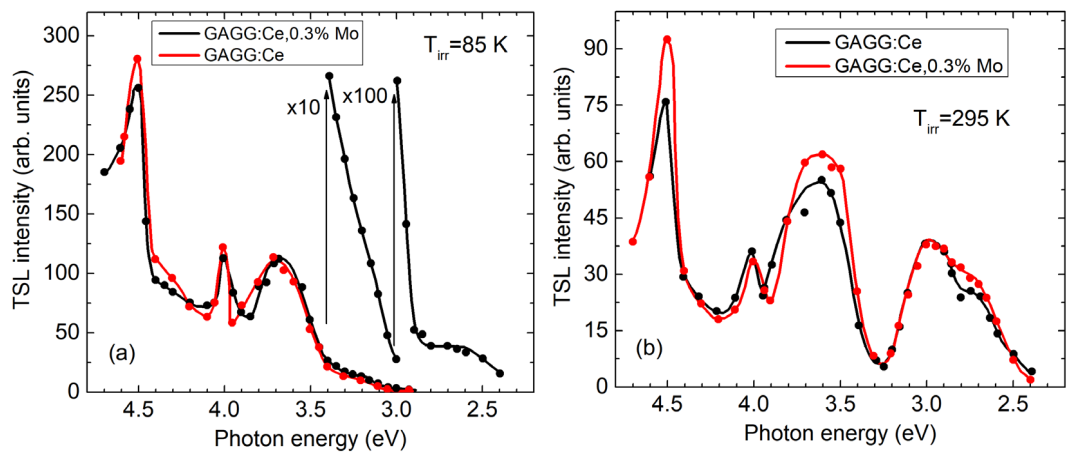

**Fig. 14.** Corrected TSL peaks creation spectra (normalized) measured after irradiation of the GAGG:Ce, 0.3% Mo (black circles) and GAGG:Ce (red circles) crystals (a) at 85 K (for the TSL peak at ≈255 K) and (b) at 295 K (for the TSL peak at ≈375 K).

### *4.5. Activation energy of TSL glow curve peaks creation*

The dependences of the intensity of the TSL glow curve peak located at 370-380 K ($I_{TSL}$) on the irradiation temperature ($T_{irr}$) presented in the $\ln(I_{TSL}) - 1/T_{irr}$ coordinates measured for the GAGG:Ce, 0.1% Mo (black filled circles), GAGG:Ce, 0.3% Mo (black empty circles), and GAGG:Ce (red filled circles) crystals are shown in Fig. 15. From these dependences, the activation energies $E_a$ of the TSL peak creation under irradiation in the $4f - 5d_1$ absorption band of $Ce^{3+}$ are determined. The same $E_a$ values (0.27 eV) are obtained for the both Mo-containing GAGG:Ce crystals (see also Table 2).

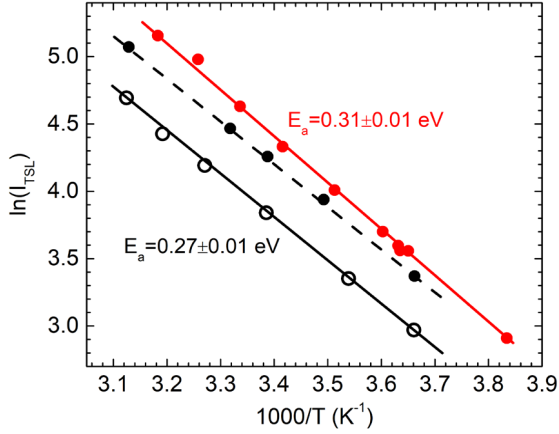

**Fig. 15.** Dependences of the TSL intensity at 375 K on the irradiation temperature $T_{irr}$ presented in the $\ln(I_{TSL})$ – $1/T_{irr}$ coordinates measured for the GAGG:Ce, 0.1% Mo crystal (filled black circles), GAGG:Ce, 0.3% Mo crystal (empty black circles), and GAGG:Ce (filled red circles). $E_{irr}$ = 2.65 eV.

The $E_a$ values are known to be very sensitive to the $Ga^{3+}$ content (x) in $Gd_3Ga_xAl_{5-x}O_{12}$:Ce crystals [26]. Namely, the smaller x is, the larger is the $E_a$ value. However, the GAGG:Ce,Mo and GAGG:Ce crystals have approximately the same $Ga^{3+}$ content (Table 1). Therefore, the difference between the $E_a$ values obtained for the GAGG:Ce (0.31 eV) and GAGG:Ce,Mo (0.27 eV) crystals allow us to conclude that the $E_a$ value is influenced by the co-doping of the GAGG:Ce crystals with $Mo^{6+}$. As the $E_q$ values in the investigated crystals are close (0.47 - 0.49 eV, see Table 2), the co-doping with Mo should not influence much the position of the $5d_1$ excited level of $Ce^{3+}$ with respect to the conduction band. Therefore, the smaller $E_a$ values can be due to larger concentration of various defects in the GAGG:Ce,Mo crystals as compared with the GAGG:Ce crystal. Indeed, the activation energy $E_a$ of the TSL peaks creation has been defined as the electron transfer energy from the excited $5d_1$ level of $Ce^{3+}$ to a defect level [26].

### 5. Discussion and final remarks

In $Ce^{3+}$ - doped compounds, three processes of the photostimulated creation of the electron and hole centers can take place:

(i) The release of electrons from the excited 5d levels of $Ce^{3+}$ takes place resulting in the optical creation of stable hole $Ce^{4+}$ center and the electrons trapped at different traps. As evident from Figs. 8 and 15, mainly thermally stimulated release of electrons appears under $4f – 5d_1$ excitation (with the activation energy $E_a$, see also Table 2). The appearance of a weak TSL under this excitation at 85 K (see Figs. 7 and 14a) can probably indicate the presence of also a small contribution of tunneling transitions into these processes. The recombination of thermally released trapped electrons with the hole $Ce^{4+}$ centers results in the appearance of different peaks at the TSL glow curve. In this case, the TSL spectrum coincides with the luminescence spectrum of $Ce^{3+}$ centers in the corresponding crystal. This well-known process considerably dominates under irradiation in the $Ce^{3+}$ - related absorption bands of GAGG:Ce.

(ii) The electron transfer from the valence band to intrinsic electron traps (e.g., oxygen vacancies, antisite ions, etc.) resulting in the optical creation of the intrinsic ($O^-$ - type, see [27]) hole centers and various electron centers. In this case, thermally stimulated electron-hole recombinations result in the appearance of the intrinsic luminescence of an exciton-like origin. This process dominates in the undoped GAGG crystal. In the $Ce^{3+}$ - doped crystals, it appears only under irradiation in the energy ranges where the $Ce^{3+}$ - related absorption is sufficiently weak.

(iii) In Gd - based compounds, an effective creation of electron and hole centers under irradiation in the absorption bands of $Gd^{3+}$ was found in [22] and investigated in detail in [21]. As the ground level of $Gd^{3+}$ is located inside the valence band of garnets [28,29,30], the excitation of a $Gd^{3+}$ ion can result not only in the intra-center emission of the $Gd^{3+}$ ion or in the immediate recombination of the released electron with a hole remained at the $Gd^{3+}$ ion accompanied with the luminescence of $Gd^{3+}$. We suggest that the irradiation of the undoped GAGG crystal in the $Gd^{3+}$ absorption bands can also result in the simultaneous release of electrons and their trapping at intrinsic crystal lattice defects and delocalization of holes from the produced $Gd^{4+}$ ions and their trapping at $O^{2-}$ ions resulting in the formation of the $O^-$ - type hole centers. Like in the process 2, the recombination of the optically created electron and hole centers results in the appearance of the intrinsic luminescence.

In the GAGG:Ce,W and GAGG:Ce,Mo crystals, owing to the presence of a huge amount of effective traps for electrons ($W^{6+}$ and $Mo^{6+}$ ions) and, consequently, a large probability of the formation of the intrinsic $O^-$ - type hole centers, the contribution of the processes 2 and 3 should be considerable. This can explain the observed differences in the characteristics of the photo- and thermally stimulated luminescence of the GAGG:Ce,W, GAGG:Ce,Mo crystals and the GAGG:Ce crystal of similar composition.

Thus, the results obtained in this work at the investigation of GAGG:Ce,W and GAGG:Ce,Mo crystals and their comparison with the characteristics of the undoped and $Ce^{3+}$ - doped W - and Mo - free GAGG crystals of similar composition allow to conclude that the co-doping with W and Mo results in their preferred stabilization in the 6+ charge state in the lattice and effective formation of the W - and Mo - related electron centers and intrinsic hole centers. The photo- and thermally stimulated electron-hole recombination results in drastic increase of the intrinsic emission contribution into the luminescence spectrum of the co-doped crystals as compared with GAGG:Ce. This effect appears in the dependence of the luminescence spectrum of the co-doped crystals on the excitation energy (Figs. 1 and 9), in the dependence of the photoluminescence thermal quenching on the preliminary irradiation conditions of the investigated sample (Figs. 3 and 11), in the afterglow decay kinetics (Figs. 4 and 12), in the TSL intensity and the shape of the TSL glow curves (Figs. 5 and 13), and in the shape of the excitation spectrum of the TSL glow curve peaks (Figs. 7 and 14). It should be noted that the influence of $Mo^{6+}$ on the characteristics of the GAGG:Ce crystal is smaller as compared with $W^{6+}$. This might be caused by the presence in the crystal lattice, besides the $Mo^{6+}$ ions, also of $Mo^{3+}$ ions (see, e.g., [16,17]), and consequently smaller concentration of the $Mo^{6+}$.

A strong influence of the pre-history of the GAGG:Ce,W and GAGG:Ce,Mo crystals on the I(T) dependence means that at the temperatures T < 400 K this dependence is determined not only by the energy distance between the excited $5d_1$ level of $Ce^{3+}$ and the bottom of the conduction band (CB) but also by the concentration and structure of defects in the investigated crystal. Thermally stimulated electron transitions from the $5d_1$ excited level of $Ce^{3+}$ to the defects levels can determine the I(T) dependence in the 300-400 K range. These transitions result in the appearance of electron centers responsible for the observed TSL glow curve peaks (with the activation energy $E_a$ around 0.25 - 0.28 eV, see Table 2). Only at higher temperatures (T > 400 K), the $5d_1$ – CB transitions become possible resulting in the photoluminescence thermal quenching with the activation energy $E_q$ around 0.46 – 0.52 eV (Table 2).

The possible reason of the irradiation-induced gradual TSL intensity reduction and changes in the I(T) dependences should be surely connected with the presence of W and Mo ions with excess positive charge in the investigated crystals as no such effect is observed in GAGG:Ce crystals. Probably, in each cycle of the crystal irradiation and heating up to 510 K, the concentration of $W^{6+}$ and $Mo^{6+}$ ions gradually decreases due to an effective trapping of electrons resulting in the creation of electron centers stable up to 510 K. The process $W^{6+} \rightarrow W^{5+} \rightarrow W^{4+} \rightarrow W^{3+}$ can reduce the concentration of various W - induced electron traps and, consequently, the number of intrinsic hole centers. This effect could explain the observed instability of the luminescence characteristics of the W - co-doped GAGG:Ce crystals.

Additionally, the co-doping of GAGG:Ce with the W and Mo ions should result in the creation of a large number of cation vacancies for the compensation of their excess positive charge. The data on compositions of the investigated crystals (Table 1) show that the Gd content is markedly lower (about 1%) than that expected for stoichiometric composition. Consequently, cation vacancy related hole centers can also be created under the UV irradiation of the co-doped crystals. These centers can also participate in the electron-hole recombination processes accompanied with the slow visible intrinsic luminescence.

Thus, in principle, the co-doping with W and Mo should negatively influence the scintillation characteristics of GAGG:Ce owing to the appearance of a huge number of slowly recombining impurity-related electron and hole centers. The positive effect of the W and Mo co-doping at their low content seems to be partly related with decrease of anti-site defects creation as was suggested in [12]. However, the main reason of improved scintillation light yield could also consist in the suppression of the electron traps related to oxygen vacancies as large positive excess charge is introduced by $W^{6+}$ or $Mo^{6+}$ ions.

**Acknowledgments**

The work was supported by the Estonian Research Council grant (PUT PRG 619), the Czech Science Foundation (project No. 20-12885S) and joint ASCR-JSPS project of Czech-Japan collaboration. Partial support of the Operational Programme Research, Development and Education financed by European Structural and Investment Funds and the Czech Ministry of Education, Youth and Sports (Project No. SOLID21 CZ.02.1.01/0.0/0.0/16_019/0000760) is also acknowledged with thanks.